\documentclass[]{aa}
\usepackage{txfonts}
\usepackage{longtable}
\usepackage{epsfig}
\usepackage{graphicx}
\begin{document}
\title{Self-consistent triaxial de Zeeuw-Carollo Models}
\author{Parijat Thakur\inst{1}
     \and Ing-Guey Jiang\inst{1} 
      \and Mousumi Das\inst{2}
      \and D.K. Chakraborty\inst{3}
      \and H.B. Ann\inst{2}}
\offprints{Dr. Parijat Thakur, 
          \email{pthakur@phys.nthu.edu.tw}}
\institute{Department of Physics and Institute of Astronomy, National Tsing-Hua 
University, Hsin-Chu 30013, Taiwan\\
\email{pthakur@phys.nthu.edu.tw}; \email{jiang@phys.nthu.edu.tw}
    \and 
     Division of Science Education, Pusan National University,
 Busan 609-735, Korea\\
    \email{mdas@pusan.ac.kr}; \email{hbann@pusan.ac.kr}
       \and 
   School of Studies in Physics, Pt. Ravishankar Shukla University, 
Raipur 492 010, India\\
   \email{ircrsu@sancharnet.in}
}
\date{Received xxxx, accepted xxxx, }
\abstract{
We use the usual method of Schwarzschild to construct 
self-consistent solutions for the triaxial  
de Zeeuw \& Carollo (1996) models with central density cusps.
ZC96 models are triaxial generalisations of 
spherical $\gamma$-models of Dehnen whose densities vary 
as $r^{-\gamma}$ near the center and $r^{-4}$ at large radii 
and hence, possess a central density core for $\gamma=0$ and 
cusps for $\gamma > 0$. 
We consider four triaxial models from ZC96, two prolate
triaxials: $(p, q) = (0.65, 0.60)$ with $\gamma = 1.0$ and $1.5$, and
two oblate triaxials: $(p, q) = (0.95, 0.60)$ with  $\gamma = 1.0$ and $1.5$.
We compute $4500$ orbits in each model for time periods of $10^{5} T_{D}$. 
We find that a large fraction of the orbits in each model are stochastic 
by means of their nonzero Liapunov exponents. The stochastic orbits in 
each model can sustain  regular shapes for $\sim 10^{3} T_{D}$ or longer, 
which suggests that they diffuse slowly through their allowed phase-space.
Except for the oblate triaxial models with $\gamma =1.0$, our attempts to construct 
self-consistent solutions employing only the regular orbits fail for the remaining
three models. However, the self-consistent solutions are found to exist 
for all models when the stochastic and regular orbits are treated in the same way 
because the mixing-time, $\sim10^{4} T_{D}$, is shorter than the integration time, 
$10^{5} T_{D}$. Moreover, the ``fully-mixed''  solutions can also be constructed  
for all models when the stochastic orbits are fully mixed at $15$ lowest energy shells. 
Thus, we conclude that the self-consistent solutions exist for our selected 
prolate and oblate triaxial models with $\gamma = 1.0$ and  $1.5$.

}
\authorrunning{Thakur P. et al.}
\titlerunning{Self-consistent triaxial de Zeeuw-Carollo Models}
\maketitle
\section{Introduction}
It is natural to assume that elliptical galaxies can be triaxial, 
such as ellipsoids (Binney $1978$). 
In this regard, Schwarzschild (1979) developed a method to explore the existence of 
triaxial ellipticals using a catalogue of numerically integrated orbits, which 
eventually became a popular tool. The triaxial models considered by 
him feature a central density core, and also show isophotal twists 
(Chakraborty \& Thakur 2000) representing the observed photometric 
properties of the majority of elliptical galaxies (Jedrzejewski 1987; Peletier et al. 1990). 
However, the ground-based (Moller, Stiavelli \& Zeilinger 1995) and 
Hubble Space Telescope (Crane et al. 1993; Jaffe et al. 1994; Ferrarese et al. 1994; 
Lauer et al. 1995; Faber et al. 1997) observations reveal that instead of
central density cores, most of the elliptical galaxies have central density cusps. 
Thus, Merritt \& Fridman $(1996)$ studied the effect of central density cusps
on triaxial configurations but their models do not show isophotal twists. 
In their study, it was found that a ``fully-mixed'' solution exists for 
a maximally triaxial model 
with a weak-cusp. The aim of this paper is to construct self-consistent 
solutions for more realistic triaxial models in the sense of representing 
both of the above-mentioned significant observed properties of 
elliptical galaxies, namely central density cusps and  isophotal twists. 
Here we used the triaxial models given by 
de Zeeuw \& Carollo (1996) (hereafter ZC96), which 
have a central density core for $\gamma=0$ and a cusp for $\gamma > 0$, 
and also show isophotal twists in their projection 
along the line-of-sight. Furthermore, it is worth to mention that 
the ZC96 triaxial models are found to be useful in 
constraining the intrinsic shapes of elliptical galaxies using 
their projected properties (Thakur \& Chakraborty 2001).

The remainder of this paper is organized as follows. In Sect. $2$, 
we present the ZC96 triaxial models. Sect. $3$ describes equations of motion 
for the ZC96 triaxial models. Sect. $4$ presents the integration 
of orbits. Sect. $5$ deals with the computation
of the Liapunov exponents. The method to construct 
the self-consistent triaxial models, is presented in  Sect. $6$. 
Sect. $7$ is devoted for the results and discussion. 
\section{Triaxial de Zeeuw-Carollo models}
We have used the triaxial potentials of ZC96, given as
\begin{eqnarray}
V(r,\theta,\phi) &=& u(r)-v(r)Y_{2}^{o}(\theta)+w(r) Y_{2}^{2}(\theta,\phi) 
\nonumber , \\
V(x,y,z) &=&  u(r)-v(r)(2z^{2}-x^{2}-y^{2})/2r^{2} + w(r)3(x^{2}-y^{2})/r^{2},
\end{eqnarray}
where $(r,\theta,\phi)$ are spherical coordinates defined such that 
$x=r\sin\theta\cos\phi$, $y=r\sin\theta\sin\phi$ and $z=r\cos\theta$, 
the functions $Y_{2}^{o}(\theta)  = \frac {3} {2} \cos^{2}\theta-\frac {1} {2}$ 
and $Y_{2}^{2}(\theta,\phi)  =  3 \sin^{2}\theta \cos2\phi$ are usual 
spherical harmonics, and $u(r)$, $v(r)$ and $w(r)$ are 
three radial functions. Here $u(r)$ is chosen to be the potential 
of the spherical $\gamma$-models of Dehnen $(1993)$, defined by
\begin{equation}
u(r) =\left\{ \begin{array}{ll} \frac{GM}{r_{o}}ln\frac{r}{r+r_{o}}, 
& \textrm{for $\gamma =2$}, \nonumber\\  
 \frac{GM}{(2-\gamma )r_{o}}[(\frac{r}{r+r_{o}})^{2-\gamma }-1], 
& \textrm{for $\gamma \ne 2$},
\end{array}\right. 
\end{equation}
where $M$ is the total mass of the model, $r_{o}$ is the scale-length and 
cusp parameter $\gamma$ can have a value in between $0\le \gamma < 3$. 
Furthermore, the functions $v(r)$ and $w(r)$ are considered as follows:
\begin{equation}
v(r)  =  - \frac {GMr_{1}{r}^{2-\gamma}} {(r+r_{2})^{4-\gamma}}, 
\qquad w(r) =  - \frac {GMr_{3}{r}^{2-\gamma}} {(r+r_{4})^{4-\gamma}},
\end{equation}
where $r_{1},...,r_{4}$ are constants. 

The associated density distribution $\rho(r, \theta, \phi)$ follows 
from Poisson's equation
\begin{eqnarray}
\rho(r, \theta, \phi) & = & f(r)-g(r) Y_{2}^{o}(\theta)+h(r) 
Y_{2}^{2}(\theta,\phi) \nonumber , \\
\rho(x,y,z) &=&  f(r)-g(r)(2z^{2}-x^{2}-y^{2})/2r^{2} + h(r)3(x^{2}-y^{2})/r^{2},
\end{eqnarray}
where $f(r)$, $g(r)$, and $h(r)$ are taken from Eq. $(2.5)$ of ZC96. 
The four ratios $r_{1}/r_{o},...,r_{4}/r_{o}$ can be expressed in terms 
of $\gamma$, and of the axial ratios of the density distribution 
at small and at large radii, respectively denoted by $(p_{o}, q_{o})$ 
and $(p_{\infty}, q_{\infty})$, where the surfaces of constant density 
are approximately ellipsoidal, i.e., $\rho \sim \rho(m^{2})$ with
$m^{2} = x^{2}+y^{2}/p^{2}+z^{2}/q^{2}$. 
ZC96 models have a central density core for $\gamma =0$, while they have cusps 
in which the density diverges as $r^{-\gamma}$ at small radii for $\gamma > 0$. 
On the other hand, the density fall off as  $r^{-4}$ at large radii.

In this paper, we fix the values of axial ratios  as $p_{o}=p_{\infty} \equiv p$ 
and $q_{o}=q_{\infty} \equiv q$ for all the calculations.

\section{Equation of motion for triaxial de Zeeuw-Carollo models}
Equation of motion is given by
\begin{equation}
\frac{d^{2}\vec{r}}{dt^{2}} = \vec{F} = -\vec{\bigtriangledown }V(x, y, z)
\end{equation}
where $\vec{F}$ is the force per unit mass, 
and $V(x, y, z)$ represents the triaxial potentials of ZC96 given in Eq. $(1)$.

The force components  $(F_{x}, F_{y}, F_{z})=(F_{1}, F_{2}, F_{3})$ 
representing the equations of motion by three scalar equations 
in the Cartesian coordinates, needed at each step of an orbit integration, 
can be calculated by the partial derivatives of $V(x, y, z)$ with respect to 
$(x, y, z) = (x_{1}, x_{2}, x_{3})$. These can be written as 
\begin{eqnarray}
F_{i} &=& -\frac{x_{i}}{r} \left [u_{D}-
\left (\frac{v_{D}}{2r^{2}}-\frac{v_{r}}{r^{3}}\right )
(2x_{3}^{2}-x_{1}^{2}-x_{2}^{2})+
A_{1} \frac{v_{r}}{r} \right.  \nonumber \\
 & &+\left. A_{2} \left(\frac{w_{D}}{r^{2}}-\frac{2w_{r}}{r^{3}}\right)
(x_{1}^{2}-x_{2}^{2})+A_{3} \frac{w_{r}}{r} \right ], 
\end{eqnarray}
where $(A_{1}, A_{2}, A_{3})$ are equal to  $(1, 3, 6)$, 
$(1, 3, -6)$, and $(-2, 3, 0)$  for $i = 1, 2, 3$, respectively. 
Furthermore,
\begin{eqnarray}    
u_{D} &=& \left(\frac{r}{r+1}\right)^{1-\gamma } 
\left[\frac{1}{r+1}- \frac{r}{(r+1)^{2}}\right], \nonumber\\
v_{D} &=& -r_{1}\left[\frac{(2-\gamma)r^{1-\gamma}}
{(r+r_{2})^{4-\gamma}}- \frac{(4-\gamma)r^{2-\gamma}}
{(r+r_{2})^{5-\gamma}}\right], \nonumber\\
w_{D} &=& -r_{3}\left[\frac{(2-\gamma)r^{1-\gamma}}
{(r+r_{4})^{4-\gamma}}- \frac{(4-\gamma)r^{2-\gamma}}
{(r+r_{4})^{5-\gamma}}\right] 
\end{eqnarray}
are the derivatives of $u_{r}$, $v_{r}$, and $w_{r}$ with respect to $r$, 
respectively. Here the expressions of $u_{r}$, $v_{r}$, and $w_{r}$ can be 
derived from those of $u(r)$, $v(r)$, and $w(r)$ given in Eqs. (2)-(3)
using $G = M = r_{o} = 1$, respectively. 
\section{Integration of orbits}
To integrate orbits, we have considered four triaxial models of ZC96, 
which are listed in Table $1$. Here the first two models are prolate 
triaxials: $(p, q) = (0.65, 0.60)$ with $\gamma = 1.0$ and $1.5$, 
while the remaining two are oblate triaxials: $(p, q) = (0.95, 0.60)$ 
with $\gamma = 1.0$ and $1.5$. In this paper, the former models will be 
called Models PT, whereas the latter models will be referred to as Models OT. 

Since the triaxial mass models of ZC96 are very centrally concentrated, 
the numerical algorithm required for integrating the orbits 
must be extremely accurate and flexible. We have used a FORTRAN 
routine, RK78, which was kindly made available by {\sf Prof. D. Pfenniger}.
This routine follows the $7/8$ order Runga-Kutta algorithm described in 
Fehlberg $(1968)$, which incorporates a variable time step in order 
to maintain a specified accuracy from one integration step to the next. 
The accuracy parameter REPS in all the integrations is chosen to be $10^{-8}$. 
Energy is typically conserved to a few parts in $10^{9}$ over $10^{5}$ 
dynamical times with this choice of REPS. 

For each of the four selected triaxial models, we have followed the scheme 
developed by Schwarzschild (1993) in assigning  initial conditions 
from the $x-z$ start-space with  $v_{x} = v_{z} = 0$ and stationary 
(equipotential) start-space with $v_{x} = v_{y} = v_{z} = 0$. 
As in Merritt \& Fridman $(1996)$, the orbits in both start-spaces 
are assigned a value from a set of $20$ energies, defined 
as the values of the potential on the $x$-axis of a set of 
$20$ ellipsoidal shells. 
The radius, energy and dynamical time ($T_{D}$) of each  ellipsoidal shell 
are given in Table $2$ for all models. 
Here energy dependent ``dynamical time'' $T_{D}$ 
is defined as the period of the $1$:$1$ resonant $x$-$y$ periodic orbit. 
For the $x-z$ start-space, a total of $150$ starting points per shell are 
calculated for all models. Moreover, the stationary 
start-space grid is defined as in Schwarzschild (1993). But, as opposed to $64$ 
starting points in Merritt \& Fridman $(1996)$, only $25$ starting 
points are calculated in each of the three sectors on an equipotential octant, 
resulting a total of $75$ starting points per shell in the stationary 
start-space. Here, as the Liapunov exponents are computed for long periods 
of $10^{5} T_{D}$ (see Sect. $5$), we reduce the number of starting points 
in order to save computing time. Thus, a total of $4500$ starting points are 
calculated for each of the selected models due to their division into a set 
of $20$ ellipsoidal shells. For each of these starting points, we have integrated 
the orbit over a time interval of $10^{5} T_{D}$.
\section{Computation of Liapunov exponents}
For detecting and quantifying stochasticity, we have followed 
Merritt \& Fridman $(1996)$ and computed 
approximations to  the six Liapunov exponents, which can be ordered 
by size as $\sigma_{1} \ge \sigma_{2} \ge ... \ge  \sigma_{6}$, 
by integrating each of 
$4500$ orbits for long periods of $10^{5} T_{D}$ using 
the Gram-Schmidt orthogonalization technique described by Benettin 
et al. (1980). In order to carry out this technique, we have used 
a FORTRAN routine, LIAMAG, which was kindly made available by 
{\sf Prof. D. Pfenniger} of Geneva Observatory group. In this routine, 
the second derivatives of the potential with respect to position are
required for determination of the evolution of the perturbed orbits  
(Udry \& Pfenniger 1988), which are given in Appendix A. 

As in Merritt \& Fridman $(1996)$, we have restricted our 
attention to the three positive Liapunov exponents, since 
$\sigma_{i} = -\sigma_{7-i}$ with $i=1,2,3$. For all models, we have 
calculated the Liapunov times ($\tau_{L}$), corresponding to 
the instability timescale between neighboring orbits, which are
found to be approximately same for all energy shells when they are 
scaled in dynamical times ($T_{D}$). Table $1$ gives these calculated values of
Liapunov time ($\tau_{L}$) in units of $T_{D}$ for all models. 
Here Models PT with $\gamma = 1.5$ appears as the most stochastic one, since 
its Liapunov time ($\tau_{L}$) is smaller compared to other selected  models. 
The sum of all three positive Liapunov exponents is defined as 
``Kolmogorov entropy'' ($h_{k} = \sum_{i=1,3} \sigma_{i}$). 
In order to distinguish regular from stochastic orbits for all models, 
unlike the case of Poon \& Merritt $(2004)$ who use
the histogram of $\sigma_{1} T_{D}$, we consider the histogram 
of $h_{k} T_{D}$ for the orbits at each energy shell from both start-spaces.
Since the histogram of $h_{k} T_{D}$ is found to have 
similar behaviours for all models, we present it in Fig. $1$ for 
the orbits at shell $13$ from the stationary start-space of 
the most stochastic model (i.e., Models PT with $\gamma = 1.5$). 
As can be seen from Fig. $1$, the separation of orbits into two groups 
representing two peaks becomes apparent as the integration time increases. 
Out of these two peaks, one is situated at narrow regions near zero showing 
the regular orbits, whereas another one is located at non-zero value with 
larger spread that decreases with the integration time, representing the stochastic orbits.
After a time interval of $\sim 10^{3} T_{D}$, $h_{k} T_{D}$ of the stochastic orbits 
at each energy shell for all models are found to approach towards common values and  
their mean values do not change very much. This suggests the followings: 
$(1)$ The stochastic orbits in all models diffuse slowly through their allowed 
phase-space, and can sustain their regular shapes for $\sim 10^{3} T_{D}$ or longer.
$(2)$ The mixing-time that is associated with diffusion through the Arnold web would be 
$\sim 10^{4} T_{D}$ (cf. Merritt \& Valluri $1996$), since it is clear from Fig. $1$ 
that there is  still a lot of mixing going on at $\sim 10^{3} T_{D}$.
Our computation of the Liapunov exponents for periods of $10^{5} T_{D}$,
which is two orders of magnitude longer than others in the literature, 
allows us to achieve these results.

At the integration time  of $10^{5} T_{D}$, 
the critical value, $h_{kc} T_{D} \approx 10^{-1.03}$, is found to separate 
the orbits into two peaks in the histograms of $h_{k} T_{D}$ at each energy 
shell from both start-spaces for all models. Thus, the orbits are
considered to be chaotic if $h_{k} T_{D} > h_{kc} T_{D}$. As can be seen 
from Fig. $2$, a large fraction of the orbits at all energy shells from 
the stationary start-space are found to be stochastic for all models. 
For each of the two selected values of $\gamma$, a larger fraction of 
the stochastic orbits are found in Models PT than in Models OT. 
Furthermore, Models PT (Models OT) with $\gamma = 1.5$ have a relatively larger 
fraction of the stochastic orbits than those of Models PT (Models OT) 
with  $\gamma = 1.0$. On the other hand, a significant fraction of the orbits 
from the $x-z$ start-space at each energy shell are also found to be stochastic 
for all models. 

\section{Construction of self-consistent models} 
Self-consistent equilibrium models can be constructed by the usual method of 
Schwarzschild $(1979)$. Here a linear superposition of orbits is sought,
each populated with an appropriate number of stars that could reproduce 
the mass of each cell. This method is formulated as 
\begin{eqnarray}
\sum_{i=1}^{M} C(i) B(i,j) = D(j) \ \ (j=1...,N) \ \ \ \textrm{with} \nonumber \\
C(i) \geq 0, \ \ \ (i=1,...,M), \ 
\end{eqnarray}
where $B(i,j)$ is the time spent by the $i^{th}$ 
orbit in the $j^{th}$ cell,  $D(j)$ is the mass of the $j^{th}$ cell,
and $C(i)$ is a weight associated with the $i^{th}$ orbit, also called 
the non-negative occupation number of that orbit. To employ Schwarzschild's method, 
all models are divided into $960$ cells 
by following the scheme set by Merritt \& Fridman $(1996)$. Furthermore,
as in Schwarzschild $(1993)$, the masses of $960$ cells and  the time spent 
by each orbit in $960$ cells are normalized to unity 
(i.e., $\sum_{j=1}^{960} D(j) = 1$ and  $\sum_{j=1}^{960} B(i, j) = 1$).
Using Lucy $(1974)$'s iterations method as formulated in Schwarzschild $(1993)$, 
we have then minimized the mean square deviation in the cell masses, i.e.,
\begin{equation}
\chi^{2} = \frac {1} {N}  \sum_{j=1}^{N} \left(D(j)-\sum_{i=1}^{M} 
C(i) B(i,j)\right)^{2} \ ,
\end{equation}
where $N=960$ and $M$ = total number of supplied orbits.
In this paper, we have used the departure from self-consistency 
to present our results, which is defined  
as $\delta = \left(\sqrt{\chi^{2}} \Big{/} {average \ mass \ per \ cell }\right)$
(cf. Merritt \& Fridman $1996$). In order to start the Lucy iterations, we first 
choose the trial values of $C(i)$=constant. 
Later, in each iteration, we compute $C(i)_{new}$ to derive $\delta$.
We continue the iterations until satisfactory convergence in $\delta$ 
is achieved, which is found to be around  the $\sim 30000$ iterations 
for all models. We repeat this procedure for an ever-larger, randomly 
selected sample of orbits and  then $\delta$ as a function of 
the number of orbits is plotted. If a self-consistent solution exists, 
$\delta$ would decrease rapidly with the number of orbits. 
\section{Results and discussion}
Our result in  Sect. $5$ supports 
the findings of previous workers (Merritt \& Fridman $1996$; 
Wachlin \& Ferraz-Mello $1998$; Siopis \& Kandrup $2000$; 
Kandrup \& Siopis $2003$; Poon \& Merritt $2004$) that a large fraction of the orbits 
from the stationary start-space for models with $\gamma>0.0$ are stochastic,
while there are significant fraction of the stochastic orbits  
in the $x-z$ start-space as well. 
Furthermore, a number of authors (Merritt \& Fridman $1996$; 
Holley-Bockelmann et al. $2001$; Holley-Bockelmann et al. $2002$; 
Poon \& Merritt $2004$) have claimed that cuspy triaxial equilibria can be 
constructed with sizable fractions of the stochastic orbits. 
This encouraged us to employ the stochastic orbits along with regular ones 
while constructing the self-consistent solutions for all models listed 
in Table $1$. In order to carry out this, we follow the method given in Sect. $6$, and 
the results are presented in Fig. $3$. Since the mixing-time, $\sim 10^{4} T_{D}$,
is shorter than the integration time, $10^{5} T_{D}$,
we attempt to construct self-consistent solutions in which
the stochastic and regular orbits are treated in the same way, allowing
each stochastic orbit to have an arbitrary occupation number. 
Thus, a total of $4500$ orbits are employed and then $\delta$ as 
a function of the number of orbits are plotted with ``circles'' in Fig. $3$.
Here $\delta$ decreases very fast with the number of orbits 
and converges well for all models. Therefore, we conclude that 
the self-consistent solutions are found to exist for all models
considered in this paper. 

Furthermore, we have run an experiment to construct the self-consistent 
solutions utilizing only the regular orbits, although there is no obvious physical 
reason why nature would host only the regular orbits. 
The results of this study are represented by ``crosses'' in Fig. $3$. 
Except for Models OT with $\gamma =1.0$, the remaining three models
do not show better convergence in $\delta$ with increasing number of provided 
orbits. So, we conclude that the regular orbits can provide a sufficient  
variety of shapes to construct the self-consistent 
solution for Models OT with $\gamma =1.0$, which has a large fraction 
of regular orbits than other selected models (see Fig. $2$).

Although we have already shown the existence of the self-consistent solutions 
by ``circles'' in Fig $3$, it would still be interesting to construct 
the ``fully-mixed'' solutions  for all models due to the reasons listed in 
Merritt \& Fridman $(1996)$. So, we, finally, attempt to construct 
the ``fully-mixed'' solutions by following Merritt \& Fridman $(1996)$. 
The results are shown in Fig. $3$ by ``square'' and ``star'' when the stochastic orbits 
are ``fully-mixed'' at $10$ and $15$ lowest energy shells, respectively. 
For each of these two cases, the convergence in $\delta$ is
found to be good, which  suggests that the ``fully-mixed'' solutions can be 
constructed for all models. Thus, we conclude that the self-consistent solutions 
exist for all models considered in this paper.
\begin{acknowledgements}
We thank the anonymous referee for useful remarks and suggestions 
that improved the present paper enormously. The computer programs 
for integrating orbits and computing Liapunov exponents 
were written by the Geneva Observatory Group and kindly made available 
to us by Prof. D. Pfenniger. The computations were done by  PC Cluster
located at Department of Physics and Institute of Astronomy, 
National Tsing-Hua University, Hsinchu, Taiwan.  
PT would like to express his sincere thanks to National Science 
Council (NSC), Taiwan, for granting postdoctoral fellowship through
grant: NSC 96-2811-M-007-006. PT and HB are also thankful to  ARCSEC 
for providing support. This study was also financially supported 
by Pusan National University in the program, Post-Doc 2004.
\end{acknowledgements}

\linespread{.8}

\begin{table*}
\caption{\small Model parameters and Liapunov time ($\tau_{L}$)}
\vskip  0.2cm
\begin{tabular}{ccccc|c}\hline \hline  
& & & & \\
Model & $p$  & $q$  & $\gamma$ & $\tau_{L}$ (in units of $T_{D}$) & Comments \\ 
& & & & &\\ \hline
& & & & &\\
&     0.65 & 0.60 & 1.0 & $\sim$21 $\pm$ 0.46  &  \\
Models PT & &     &  & & Prolate Triaxial  \\
   &   0.65 & 0.60 & 1.5 & $\sim$13 $\pm$ 0.33  & \\ 
   & & & & & \\ \hline  
   & & & & \\
 &   0.95 & 0.60 & 1.0 &  $\sim$25 $\pm$ 0.50  &\\
Models OT & & & &  & Oblate Triaxial \\
  &   0.95 & 0.60 & 1.5 &  $\sim$17 $\pm$ 0.40 & \\
 & & & & & \\ \hline
\end{tabular}
\end{table*}
\begin{table*}
 \centering
\begin{minipage}{310mm}
\caption{\small Shell parameters for Models PT and OT}
\vskip  0.2cm
\begin{tabular}{cccccccccclll}  \hline \hline \\
{Shell}  & \multicolumn{2}{c} {\underline{Radius \footnote{On x-axis}}} & 
\multicolumn{2}{c} {\underline{Energy for Models PT}} &   \multicolumn{2}{c} 
{\underline{$T_{D}$ for Models PT}} & \multicolumn{2}{c} {\underline{Energy for Models OT}} &
 \multicolumn{2}{c} {\underline{$T_{D}$ for Models OT}} \\
      &  $\gamma=1.0$ & $\gamma=1.5$  & $\gamma=1.0$ & $\gamma=1.5$ 
& $\gamma=1.0$ & $\gamma=1.5$  & $\gamma=1.0$ & $\gamma=1.5$ & $\gamma=1.0$ 
& $\gamma=1.5$ \\ \\ \hline \\
1 & 0.2791 & 0.1512 & -0.8103 & -1.3203 & 2.916 & 1.059  & -0.7963 & -1.2974 
& 3.204  & 1.156\\ 
2 & 0.4464 & 0.2635 & -0.7265 & -1.1378 & 3.980 & 1.668  & -0.7089 & -1.1116 
& 4.356  & 1.812\\
3 & 0.6076 & 0.3760 & -0.6600 & -1.0073 & 4.956& 2.260   & -0.6407 & -0.9800 
& 5.412 & 2.452\\ 
4 & 0.7744 & 0.4949 & -0.6023 & -0.9017 & 5.964& 2.876   & -0.5824 & -0.8744 
& 6.492 & 3.116\\ 
5 & 0.9530 & 0.6238 & -0.5503 & -0.8112 & 7.044 & 3.548  & -0.5304 & -0.7845 
& 7.644 & 3.836\\
6 & 1.1483 & 0.7662 & -0.5024 & -0.7310 & 8.244 & 4.299  & -0.4830 & -0.7054 
& 8.916 & 4.636\\
7 & 1.3660 & 0.9259 & -0.4575 & -0.6584 & 9.604 & 5.156  & -0.4390 & -0.6343 
& 10.357 & 5.548\\
8 & 1.6124 & 1.1075 & -0.4151 & -0.5920 & 11.173 & 6.156 & -0.3978 & -0.5694 
& 12.013 & 6.604\\
9 & 1.8956 & 1.3172 & -0.3747 & -0.5298 & 13.030 & 7.332 & -0.3589 & -0.5094 
& 13.958 & 7.852\\
10 & 2.2265 & 1.5628 & -0.3362 & -0.4718 & 15.263 & 8.764  & -0.3220 & -0.4535 
& 16.303 & 9.356\\
11 & 2.6199 & 1.8555 & -0.2993 & -0.4171 & 18.014 & 10.533  & -0.2867 & -0.4011 
& 19.181 & 11.205\\
12 & 3.0972 & 2.2114 & -0.2637 & -0.3654 & 21.499 & 12.774 & -0.2529 & -0.3516 
& 22.803 & 13.542\\
13 & 3.6903 & 2.6544 & -0.2296 & -0.3164 & 26.033 & 15.703 & -0.2205 & -0.3048 
& 27.505 & 16.575\\
14 & 4.4495 & 3.2220 & -0.1967 & -0.2697 & 32.174 & 19.661  & -0.1893 & -0.2603 
& 33.845 & 20.668\\
15 & 5.4580 & 3.9767 & -0.1650 & -0.2252 & 40.877 & 25.282 & -0.1593 & -0.2179 
& 42.800 & 26.457\\\
16 & 6.8661 & 5.0312 & -0.1345 & -0.1829 & 54.026 & 33.781  & -0.1304 & -0.1775 
& 56.265 & 35.164\\
17 & 8.9736 & 6.6103 & -0.1052 & -0.1424 & 75.726 & 47.805 & -0.1024 & -0.1388 
& 78.397 & 49.468 \\
18 & 12.4807 & 9.2393 & -0.0771 & -0.1040 & 116.652 & 74.287 & -0.0754 & -0.1018 
& 119.946 & 76.358 \\
19 & 19.4875 & 14.4930 & -0.0502 & -0.0674 & 214.727 & 137.42  & -0.0494 & -0.0664 
& 219.139 & 140.18\\
20 & 40.4939 & 30.2466 & -0.0245 & -0.0327 & 605.260 & 392.30 & -0.0243 & -0.0325 
& 611.661 & 396.64 \\
\\ \hline 
\end{tabular}
\end{minipage}
\end{table*}
\begin{figure}
\resizebox{\hsize}{6cm}{\includegraphics[scale = 0.6, trim = 150 215 100 100, clip]{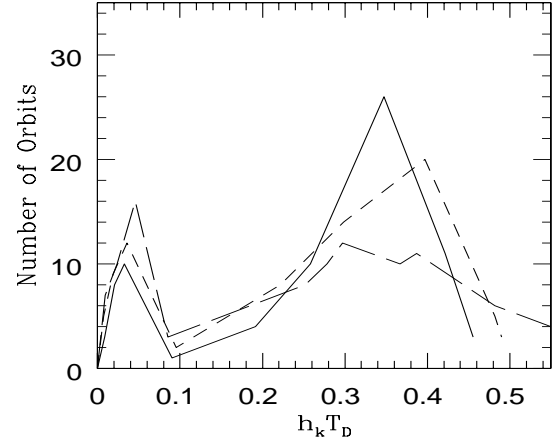}}
\caption{Histogram of ${h_{k} T_{D}}$ for the orbits at shell $13$ from
the stationary start-space of Models PT with $\gamma = 1.5$, which is 
the most stochastic model. Long-dash line: $t = 10^{3} T_{D}$;
short dash line: $t = 10^{4} T_{D}$; solid line: $t = 10^{5} T_{D}$.}
\end{figure}
\begin{figure}
\resizebox{\hsize}{10cm}{\includegraphics[scale = 1.0, trim = 0 8 0 0, clip]{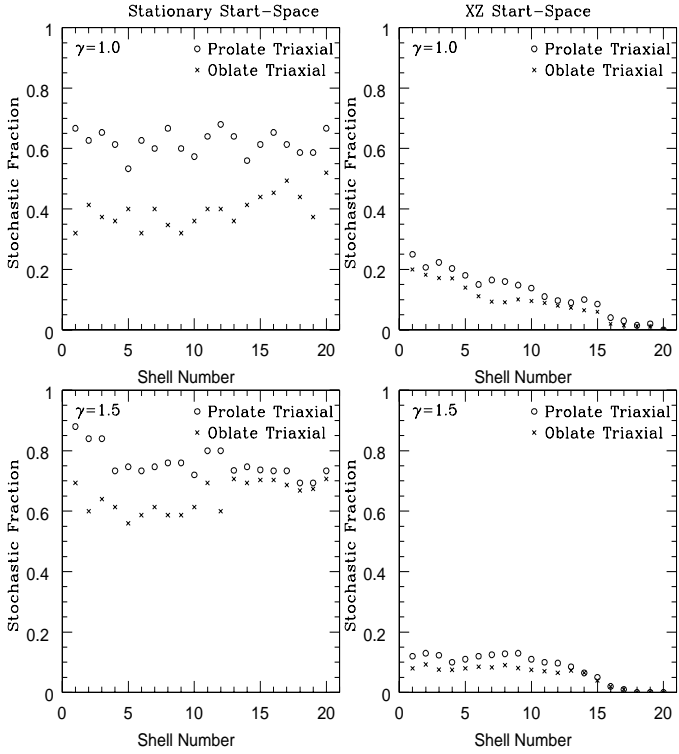}}
\caption{Fraction of the stochastic orbits per shell in the orbit libraries. 
The left panels are for the stationary start-space, whereas the right ones are for 
the $x-z$ start-space. The cusp parameter $\gamma$ is given in top left corner 
of each panel. Circles: Models PT; crosses: Models OT.}
\end{figure}
\begin{figure}
\resizebox{\hsize}{10cm}{\includegraphics[scale = 1.0, trim = 15 8 0 0, clip]{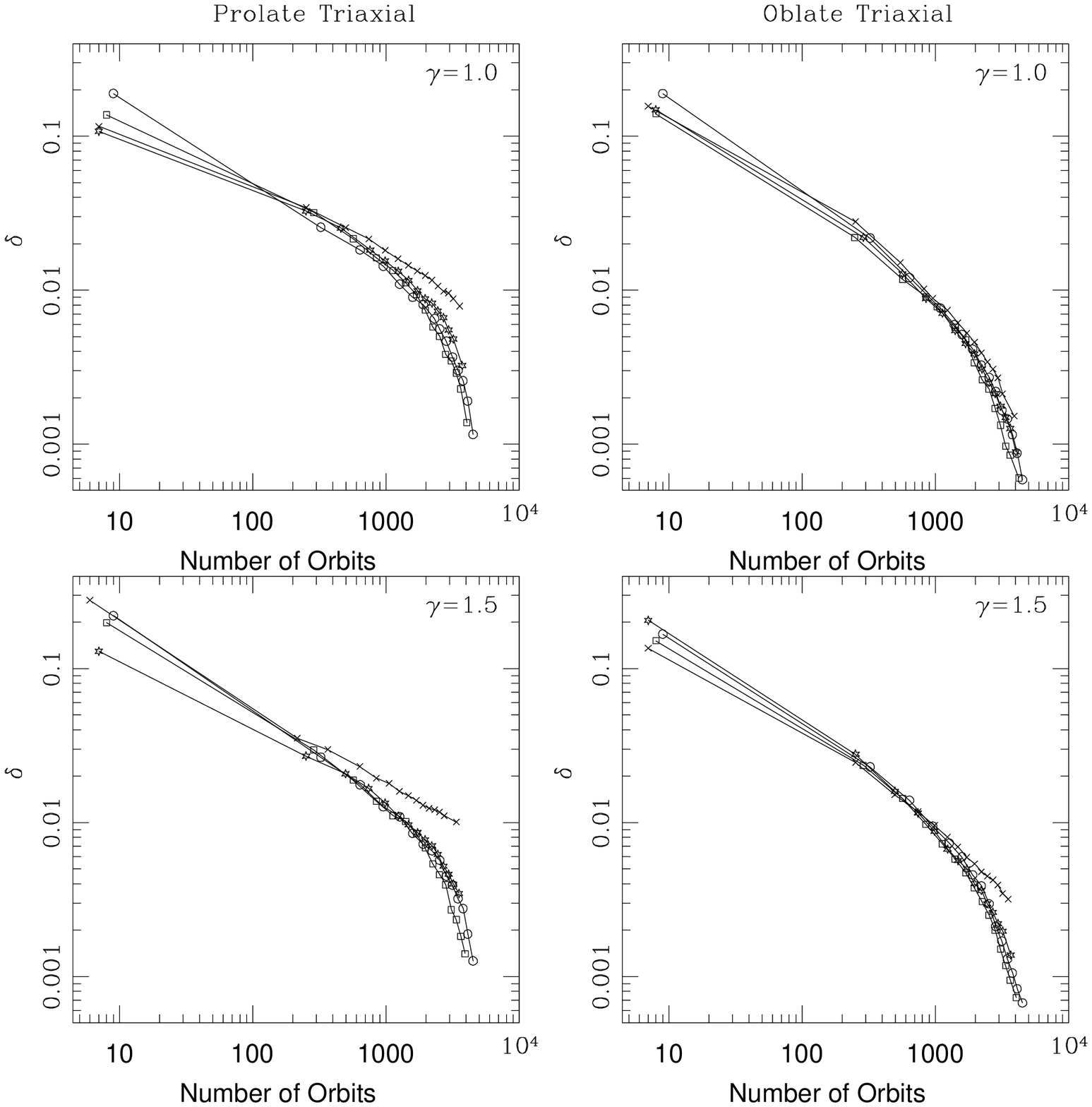}}
\caption{Departure from self-consistency $(\delta)$ as a function of
the number of orbits supplied to the minimization routine. 
Left panels: Models PT with $\gamma = (1.0, 1.5)$; 
right panels: Models OT with $\gamma = (1.0, 1.5)$. 
Circles: all orbits; crosses: regular orbits only; 
squares: the stochastic orbits are fully mixed at $10$ lowest 
energy shells; stars: the stochastic orbits are fully mixed at 
$15$ lowest energy shells.}
\end{figure}
\appendix
\section{The force derivatives} 
Here we present the derivatives of the gravitational forces corresponding 
to the ZC96 models, which are 
calculated by the gradient of the three components of forces, i.e.,  
$F_{i}$ with $i = 1, 2, 3$, in the Cartesian coordinates. The  derivatives, 
$F_{ii} = \frac{\partial F_{i}}{\partial{x_{i}}}$, can be written as
\begin{eqnarray}
F_{ii} &=& -\left [u_{DD} \ \left(\frac{x_{i}}{r}\right)^2 + u_{D} \ T_{7} - 
\left(\frac{x_{i}}{r}\right)^2 T_{1} \ T_{5}  - T_{3} \ \left(-\frac {2 x_{i}^2} {r} 
+ T_{1} \ T_{7} \right) \right.  \nonumber \\
& &+ \left. \frac{v_{r}}{r^{2}} + \frac {2 x_{i}^2} {r} \  T_{3}
+ 3\ \left(\frac{x_{i}}{r}\right)^2 T_{2} \ T_{6} 
+ 3 \ T_{4} \ \left(\frac {2 x_{i}^2} {r} + T_{2} \ T_{7}\right) \right.  \nonumber \\
& &+ \left. 6 \ \left(\frac{w_{r}} {r^2} + \frac{x^2} {r} \ T_{4} \right) \right], \  
\textrm{for $i = 1, 2, 3$}.
\end{eqnarray}
Furthermore, the cross derivatives, 
$F_{ij} = \frac{\partial F_{j}}{\partial{x_{i}}}$ with $i \ne j$, are
defined as
\begin{eqnarray}
F_{ij} &=& -\frac{x_{i} x_{j}}{r} \left[\frac{u_{DD}}{r} - \frac{u_{D}}{r^2} \ - 
 \frac{T_{1}}{r} \ T_{5} - T_{3} \ \left (-2 - \frac {T_{1}} {r^2}\right)
+  2 \ T_{3} \right.  \nonumber \\
& &+ \left. \frac{3 \ T_{2}}{r} \ T_{6} + \ 3 \left(2 - \frac {T_{2}} {r^2}\right) 
\ T_{4} \ - \ 6  \ T_{4} \right],  \ \textrm{for $i \& j = 1, 2, 3$}.
\end{eqnarray}
In Eqs. (A.1)-(A.2),  the terms $T_{1}$, $T_{2}$, $T_{3}$, 
$T_{4}$, $T_{5}$, $T_{6}$, and $T_{7}$ have following forms:
$T_{1} = \left(2x_{3}^2 - x_{1}^2 - x_{2}^2\right)$, $T_{2} = \left(x_{1}^2 - x_{2}^2\right)$,
$T_{3} = \left(\frac{v_{D}} {2r^2} - \frac{v_{r}} {r^3}\right)$,
$T_{4} = \left(\frac{w_{D}} {r^2} - \frac{2w_{r}} {r^3}\right)$,
$T_{5} = \left(\frac{v_{DD}} {2r^2} -  \frac{2v_{D}} {r^3} + \frac{3v_{r}} {r^4}\right)$
$T_{6} = \left(\frac{w_{DD}} {r^2} -  \frac{4w_{D}} {r^3} + \frac{6w_{r}} {r^4}\right)$, and
$T_{7} = \left(\frac {1} {r} - \frac {x_{1}^2} {r^3}\right)$.
Here $u_{DD}$, $v_{DD}$, and  $w_{DD}$ are given by
\begin{eqnarray}
u_{DD} &=& \left(\frac{r} {r+1}\right)^{2-\gamma} \left(\frac {1} {r+1} - 
\frac {1} {r} \right) \left[\frac {2 } {r+1} + (1-\gamma) 
\left(\frac {1} {r+1} - \frac {1} {r} \right)\right], \nonumber \\
v_{DD} &=& - \frac {r_{1} \ r^{-\gamma}} {(r+r_{2})^{4-\gamma}}  
\left[ (1-\gamma) (2-\gamma) - 2 (2-\gamma) (4-\gamma) \frac {r} {(r+r_{2})} \right.  \nonumber \\
& & +  \left. (4-\gamma) (5-\gamma) \frac {r^2} {(r+r_{2})^2} \right], \nonumber \\
w_{DD} &=& - \frac {r_{3} \ r^{-\gamma}} {(r+r_{4})^{4-\gamma}} 
\left[ (1-\gamma) (2-\gamma) - 2 (2-\gamma) (4-\gamma) \frac {r} {(r+r_{4})} \right.  \nonumber \\
& & +   \left. (4-\gamma) (5-\gamma) \frac {r^2} {(r+r_{4})^2} \right], 
\end{eqnarray}
which represent the derivatives of $u_{D}$, $v_{D}$, and  $w_{D}$ with respect 
to $r$, respectively.

\small
\begin{thebibliography}{}
\bibitem{} Benettin, G., Galgani, L., Giorgilli, A., \& Strelcyn, J.-M. 
1980, Meccanica, 15,21
\bibitem{} Binney, J.J. 1978, Comments on Astrophysics, 8, 27
\bibitem{} Chakraborty, D.K., \& Thakur, P. 2000, MNRAS, 318, 1273
\bibitem{} Crane, P., et al. 1993, AJ, 106, 1371 
\bibitem{} Dehnen, W. 1993, MNRAS, 265, 250
\bibitem{} de Zeeuw, P.T., \& Carollo, C.M. 1996, MNRAS, 281, 1333.(ZC96)
\bibitem{} Faber, S. M., et al. 1997, AJ, 114, 1771
\bibitem{} Fehlberg, E. 1968, NASA, Tech. Rep. TR R-287
\bibitem{} Ferrarese, L., van den Bosch, F.C., Ford, H.C., Jaffe, W., 
\& O'Connell, R.W. 1994, AJ, 108, 1598
\bibitem{} Holley-Bockelmann, K., Mihos, J.C., Sigurdsson, S., \& 
Hernquist, L. 2001, ApJ, 549, 862
\bibitem{} Holley-Bockelmann, K., Mihos, J.C., Sigurdsson, S., \& 
Hernquist, L. 2002, ApJ, 567, 817
\bibitem{} Jaffe, W., Ford, H.C.,  O'Connell, R.W.,  
van den Bosch, F.C., \& Ferrarese, L. 1994, AJ, 108, 1567
\bibitem{} Jedrzejewski, R.I., in de Zeeuw P.T. 1987, eds, Proc. 
IAU Symp. Vol 127, Structure and Dynamics of Elliptical Galaxies, 
Dordrecht: Reidel, p. 37
\bibitem{} Kandrup, H.E., \& Siopis, C. 2003, MNRAS, 345, 727 
\bibitem{} Lauer, T. et al. 1995, AJ, 110, 2622
\bibitem{} Lucy, L.B. 1974, AJ, 79, 745
\bibitem{} Merritt, D., \& Fridman, T. 1996 ApJ, 460, 136
\bibitem{} Merritt, D., \& Valluri, M. 1996, ApJ, 471, 82
\bibitem{} Moller, P., Staivelli, M., \& Zeilinger, W.W. 1995, MNRAS, 276, 979
\bibitem{} Peletier, R.F. et al. 1990, AJ, 100, 1091
\bibitem{} Poon, M.Y., \& Merritt, D. 2004, ApJ, 606, 774 
\bibitem{} Schwarzschild, M. 1979, ApJ, 232, 236
\bibitem{} Schwarzschild, M. 1993, ApJ, 409, 563 
\bibitem{} Siopis, C., \& Kandrup, H.E. 2000, MNRAS, 319, 43
\bibitem{} Thakur, P., \& Chakraborty, D.K. 2001, MNRAS, 328, 330
\bibitem{} Udry, S., \& Pfenniger, D. 1988, A\&A, 198, 135
\bibitem{} Wachlin, F.C., \& Ferraz-Mello, S. 1998, MNRAS, 298, 22
\end{thebibliography}
\end{document}